\documentclass[12pt]{article}\usepackage[hyperfootnotes=false]{hyperref}
   \usepackage{epsfig}
      \usepackage{amsmath}
      \usepackage{amssymb}
  \usepackage{graphicx}
  \usepackage{color}
  \newlength{\abstractwidth}
  \renewcommand{\thefootnote}{\fnsymbol{footnote}}
  \renewcommand{\thanks}[1]{\footnote{#1}} % Use this for footnotes
  \newcommand{\starttext}{
  \setcounter{footnote}{0}
  \renewcommand{\thefootnote}{\arabic{footnote}}}
  \renewcommand{\theequation}{\thesection.\arabic{equation}}
  \newcommand{\be}{\begin{equation}}
  \newcommand{\bea}{\begin{eqnarray}}
  \newcommand{\eea}{\end{eqnarray}}
  \newcommand{\beq}{\begin{equation}}
  \newcommand{\ee}{\end{equation}}
  \newcommand{\eeq}{\end{equation}}

  \def\ba{\begin{eqnarray}}
  \def\ea{\end{eqnarray}}

  \def\12{{1 \over 2}}

 \def\simleq{\; \raise0.3ex\hbox{$<$\kern-0.75em
      \raise-1.1ex\hbox{$\sim$}}\; }
 \def\simgeq{\; \raise0.3ex\hbox{$>$\kern-0.75em
      \raise-1.1ex\hbox{$\sim$}}\; }

\def\O2{\Omega_2}

\def\bi{\begin{itemize}}
  \def\ei{\end{itemize}}

\def\sc{\setcounter{equation}{0}}

\def\W{$\Omega$}
\def\W'{$\Omega$}

\def\V{\Omega}
\def\V'{\Omega}

\def\nref#1{(\ref{#1})}

\begin{document}
  \renewcommand{\theequation}{\thesection.\arabic{equation}}

\begin{titlepage}
  \rightline{}
  \bigskip

  \bigskip\bigskip\bigskip\bigskip

  \bigskip

\centerline{\Large \bf { ~~ Exact results for the entanglement entropy \\
 }}
\centerline{\Large \bf { and the energy radiated by a quark
}}
    \bigskip

  \begin{center}

 \bf { Aitor Lewkowycz$^1$ and Juan Maldacena$^2$}
  \bigskip \rm
\bigskip

$^1$ Jadwin Hall,  Princeton University, Princeton, NJ 08544, USA
\\ ~~
\\
$^2$ Institute for Advanced Study,  Princeton, NJ 08540, USA  \\
\rm

\bigskip
\bigskip

\vspace{2cm}
  \end{center}

  \bigskip\bigskip

 \bigskip\bigskip
  \begin{abstract}

We consider a spherical region with a heavy quark in the middle. We compute the
extra entanglement entropy due to the presence of a heavy quark both in ${\cal N}=4 $ Super Yang Mills
and in the ${\cal N}=6$ Chern-Simons matter theory (ABJM).
This is done by relating the computation to the expectation value of a circular Wilson loop and a
stress tensor insertion.

We also give an exact expression for the Bremsstrahlung function that determines the energy
radiated by a quark in the ABJM theory.

 \medskip
  \noindent
  \end{abstract}

  \end{titlepage}

    \starttext \baselineskip=17.63pt \setcounter{footnote}{0}
  \tableofcontents

  \sc

\section{Introduction}

\begin{figure}[h!]
\begin{center}
\vspace{5mm}
\includegraphics[scale=0.85]{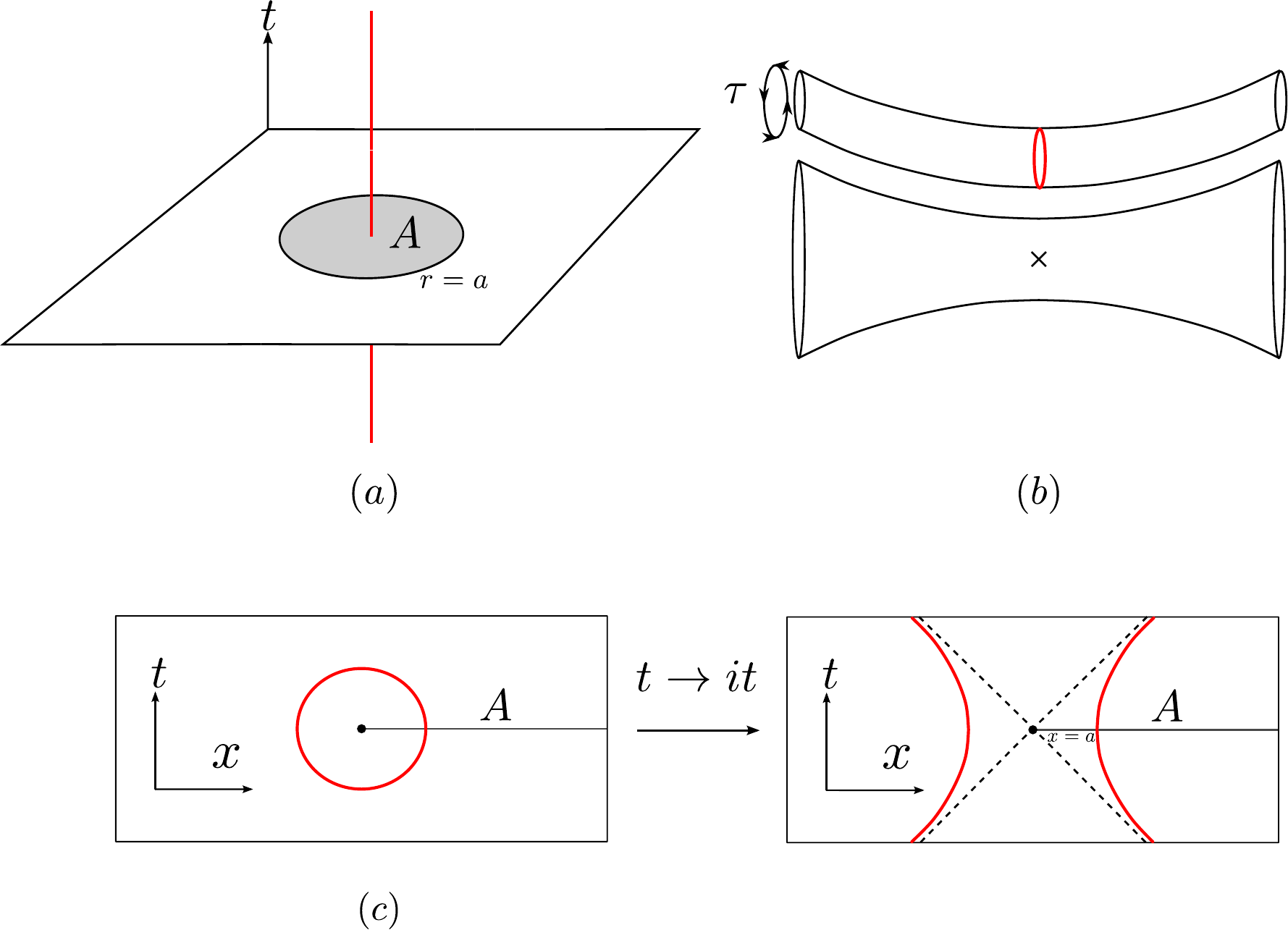}
\vspace{5mm}
\caption{
$(a)$ is the entanglement
entropy of a disk in the presence of a Wilson line$^{\ref*{FootNoteForFigureCaption}}$,  $(b)$ is the thermal entropy in thermal hyperbolic space in the presence of a
Polyakov loop and $(c)$ is the entropy of a plane in the presence of a circular loop, which can also be interpreted as the entropy induced by  a quark/anti-quark pair  undergoing hyperbolic motion .}
\label{loop}
\end{center}
\end{figure}

We compute the entanglement entropy of a spherical region that contains an external heavy quark (or Wilson line) at its center, see figure \ref{loop}(a).
We compute the additional entanglement entropy relative to the one present in the vacuum. This is a UV finite quantity.
In a conformal field theory this computation can be done if one knows the result for the circular Wilson loop
as well as the one point function of the stress tensor in the presence of the circular Wilson loop.
The reason is simple, this problem can be mapped to the computation of the ordinary entropy for a thermal
field theory on hyperbolic space at inverse temperature $\beta = 2 \pi$ \cite{Casini:2011kv}.
To compute the entropy we need to know the
free energy as well as its first derivative with respect to the temperature. The latter can be computed by
slightly changing the size of the thermal circle in
 the euclidean geometry, which is of the form $S^1 \times H_{d-1}$. This is achieved
by an insertion of the stress tensor. Both of these can be computed at inverse temperature $\beta = 2 \pi$ since
this space is   conformal to ordinary flat space. In flat space the one point function of the stress tensor in the
presence of a Wilson loop is fixed by conformal symmetry up to an overall coefficient.

Then the additional entanglement entropy due to the Wilson loop has the form
\begin{eqnarray}
 S_{W}=(1-n\partial_n) \log \langle W  \rangle|_{n=1}=\log \langle W \rangle+\int \langle T_{\tau \tau} \rangle_W
\end{eqnarray}

The simplest example is a pure Chern Simons theory. Here $T=0$ and the entanglement entropy is just given by
the Wilson loop expectation value \cite{Kitaev:2005dm,Levin:2006zz}.

In certain supersymmetric field theories one has exact methods for reducing the computation of the Wilson loop to a
certain matrix integral \cite{Pestun:2007rz,Kapustin:2009kz}.
It is also possible to compute the one point function of the stress tensor in the presence of the
Wilson loop. We will provide a precise way to relate these computations to the entanglement entropy in question.
In three dimensional theories, \cite{Nishioka:2013haa}
 have defined a certain supersymmetric R\'{e}nyi entropy which coincides with the
ordinary entanglement entropy as we take the replica number  $n\to 1$. We can also apply their method to this computation and
the answer is related to the computation of a Wilson loop in the $b$-deformed theory \cite{Hama:2011ea}.

The results are as follows. For ${\cal N}=4$ super Yang Mills we obtain
\begin{equation} \label{nfourres}
S_W = ( 1 - \frac{ 4}{ 3 } \lambda \partial_{\lambda} ) \log \langle W_{\circ} \rangle
\end{equation}
where $\langle W_{\circ} \rangle$ is the expectation value of the circular Wilson loop
 in the appropriate representation.
%%%%  FOOTNOTE
\addtocounter{footnote}{1}\footnotetext{  Note that we will always be considering the circular loop, so we actually have another line at infinity. \label{FootNoteForFigureCaption}}
%%%% FooTNOTE
For ${\cal N}\ge 2$ theories in three dimensions we obtain
\begin{equation} \label{thredim}
S_W = ( 1 - \frac{ 1 }{2} \partial_b ) \log \langle W_b \rangle |_{b=1}
\end{equation}
where $\langle W_b \rangle $ is the circular Wilson loop expectation value on the $b$-deformed sphere.

\begin{figure}[h!]
\begin{center}
\vspace{5mm}
\includegraphics[scale=0.8]{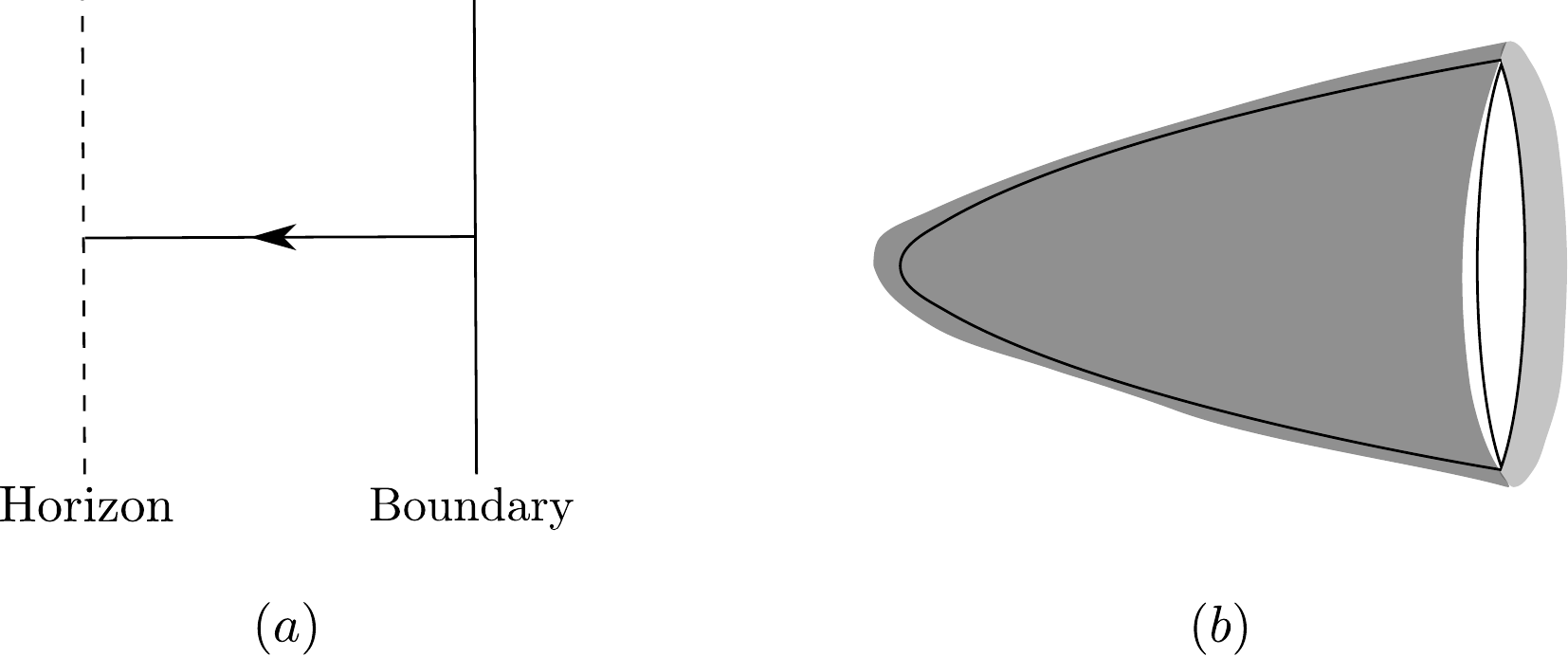}
\vspace{5mm}
\caption{ $(a)$ String coming from the boundary and ending at the horizon. $(b)$ In Euclidean space we have a worlsheet wrapping the radial and time directions with a disk topology. We can use it to compute the entropy.
}
\label{stringhor}
\end{center}
\end{figure}

In theories with gravity duals this entropy computation can be related to the entropy of a string that ends on
the horizon of a hyperbolic black hole. As noted in \cite{Jensen:2013ora,Jensen:2013ora3,Karch:2008uy,Sonner:2013mba,Chernicoff:2013iga},
when a string ends on the horizon of a black hole there is
an extra contribution to the entropy that arises from the string. This contribution has two pieces. One is related
to a term of the form $-\log g_s$ that comes from the fact that the topology of the Euclidean worldsheet is a disk, see figure \ref{stringhor}$(b)$.
This is related to a factor of $N$ that appears in the computation of a Wilson loop.
In addition,  we have a  term that
depends on the string tension. The simplest contribution
comes from the area of a  euclidean string worldsheet wraps around
  the Euclidean black hole cigar topology.
The  exact results discussed in this paper interpolate between the weak coupling answers and the results previously
obtained using classical strings.

The one point function of the stress tensor in the presence of a Wilson loop is also expected to determine the amount of
radiation produced by a moving quark. The reason is that the   circular Wilson loop can be mapped to two accelerated quarks and
we can integrate the stress tensor flowing outside the Rindler regions. This computation is slightly subtle because the
particle accelerates forever and it is difficult to separate the energy that is radiated from the increasingly
boosted value  of the self energy (see \cite{Fulton:1960,Boulware:1979qj} and references therein).
However,  for certain supersymmetric theories we propose a method for removing the
self energy contribution by using the
 existence of a vacuum expectation value for an operator $O$ of dimension $d-2$ in the presence of the
Wilson loop. In those cases, we can obtain an expression for the coefficient of the radiated energy, called the Bremsstrahlung
function, in terms of the stress tensor expectation value, see equation \nref{Bfunction}.
Using the results in \cite{Marino:2009jd,Klemm:2012ii} we give an expression for $B$ in the planar ABJM theory \cite{Aharony:2008ug}.
  If it were possible to also derive an expression for this function using integrability (as it was done
for ${\cal N}=4 $ super Yang Mills in \cite{Correa:2012hh,Drukker:2012de}),
 then by comparing the two answers one could compute a non-trivial function of the
t'Hooft coupling that appears in all the ABJM results obtained via integrability \cite{Gromov:2008qe}.

The structure of the paper is as follows. In section 2, we explain the configuration that we are going to consider. In section 3, we compute the entropy for  a free scalar and a free vector. In section 4, we compute the entanglement entropy for ${\cal N}=4$ SYM and check the expression at weak and strong coupling. In section 5, we study $3$d Chern-Simons-matter theories. In particular we obtain the entropy for Wilson loops in ABJM. In section 6, we construct a Bremsstrahlung function for superconformal field theories and propose a $B$ function for ABJM. In section 7, we briefly make some comments about the entanglement entropy from the worldsheet perspective.

\section{The configuration}

Let us present the configuration that  we are considering.
%For  simplicity, let's focus in $4d$ in this section.
%Most of the discussion generalizes trivially to generic dimensions.

Consider a circular Wilson loop in the $x,t$ directions, centered at $0$. See figure \ref{loop}(c).
  We are going to compute its entanglement entropy. From the Lorentzian perspective, this is the entanglement
between two quarks undergoing uniformly accelerated motion. That is, if the quarks are separated a distance $2a$ in the direction $x$ at $t=0$, our entangling surface is $x=0$. If we go to ``replica trick'' coordinates,
$\tau $, $r=\sqrt{x^2+t^2}$, then our entangling surface is $r=0$ and the Wilson loop is the circle $r=a$ (see figure \ref{loop}(c) ). The metric is
\begin{equation} \label{MetricD}
 ds^2=dr^2+r^2 d\tau^2+dy^2+y^2 d\Omega_{d-3}^2
\end{equation}
Proceeding as in \cite{Casini:2011kv}, the computation of the entanglement entropy is equal to the thermal entropy in the space $d s_{ S^1 \times H_{d-1}}^2=\dfrac{1}{r^2} ds^2$.

We can now rewrite the metric in hyperbolic coordinates
\begin{eqnarray}
 ds_{S^1 \times H_{d-1}} ^2&=&d \tau^2+\dfrac{dr^2+dy^2+y^2 d \Omega_{d-3}^2}{r^2} \nonumber \\
  &= & d \tau^2+d\rho^2+\sinh^2 \rho( d\theta^2+\sin^2\theta d\Omega_{d-3}^2)  \label{eq:myersgeo}
 \end{eqnarray}
In terms of these coordinates the Wilson loop is simply a Polyakov loop sitting at $\rho=0$, wrapped
along the $\tau$ direction.
The replica trick now is as simple as changing the radius of the $S^1$, $\tau \sim \tau+ \beta$, the original geometry has $\beta=2\pi$. See Appendix \ref{sec:coord} for a description of various coordinate systems.

Using the geometry \nref{eq:myersgeo}, we would like to compute the thermal entropy $S_W=(1-\beta \partial_{\beta} ) \log W$. If change to coordinates $\sigma=\beta \tau$ (so $\sigma \sim \sigma + 2\pi$), we can evaluate the derivative in terms of the stress tensor

\begin{equation} \label{OneTpoint}
 \beta \partial_{\beta}  W|_{\beta = 2\pi}= \int_{S^1 \times H^{d-1}} -\beta \langle \dfrac{\partial (\sqrt{g} {\cal L} ) }{\partial g^{\mu \nu}} \partial_{\beta} g^{\mu \nu}\rangle_W=\int 2\beta^{-2} \langle \dfrac{\partial (\sqrt{g} {\cal L} )}{\partial g^{\sigma \sigma}}\rangle_W =-\int \sqrt{g} \langle T_{\tau \tau} \rangle_W
\end{equation}

Where we used the usual stress tensor  $\sqrt{g} T_{\mu \nu}=-2 \dfrac{\partial (\sqrt{g} {\cal L} ) }{\partial g^{\mu \nu}} $.
 Of course this is just saying that the operator that changes $\beta$ is $T_{\tau \tau}$\footnote{ There could also be a contribution from the change in the metric in the part of the Wilson loop involving the scalar. This leads to the insertion of
 an operator along the loop whose one point function  vanishes by conformal symmetry.   }.
So we have to compute

\begin{equation} \label{EntropyOne}
 S_W=\log W+\int \langle T_{\tau \tau} \rangle_W
\end{equation}

{\bf A subtlety in the definition}

When we compute the entanglement entropy on a state which is not the vacuum we encounter a small finite ambiguity  related
to the precise procedure for defining the entropy. Namely, when we consider the replica trick we need to consider the
partition function of a theory on a conical space. Depending on how we regularize the cone we can get additional finite
contributions that depend on the state \cite{Faulkner:2013ana,Kabat:1995eq,Solodukhin:2011gn,Donnelly:2012st}. These arise because of the conformal coupling to the scalars which involve
$\int R \phi^2 $. Such terms can give rise to delta function contributions at the tip of the cone. We then get extra
 terms that involve the expectation value of $\phi^2$ at the entangling surface, which can have a finite part.
Such  extra terms are harmless but should be clearly specified to make sense of the computation. It is interesting to note that relative entropy  does  not suffer from this ambiguity. The ambiguity in 
the change of entanglement entropy is canceled with the ambiguity in defining the modular Hamiltonian \footnote{We thank Horacio Casini for pointing this to us.}.    A more detailed discussion can be found in appendix \ref{sec:wald}.

\subsection{Stress tensor for CFT's in the presence of a Wilson line}
\label{sec:stress}

Conformal invariance, tracelessness and conservation are strong constraints for the stress energy tensor.
In the presence of a Wilson line we have enough unbroken symmetry to completely fix the functional
dependence of the stress tensor one point function. Only the overall constant remains to be computed.
For example, if we use hyperbolic coordinates \nref{eq:myersgeo}
, the stress tensor cannot depend on $\tau$ and it has to be invariant under $\tau \to - \tau$.
 This, plus rotational symmetry constrains it to be diagonal. In addition, it should be traceless and conserved.
 Using the metric \nref{eq:myersgeo} and setting the Wilson loop along the $\tau $ direction at $\rho=0$ we find
 the following expectation value for the stress tensor
\begin{equation}
\langle T_{\mu \nu} \rangle_W dx^{\mu} dx^{\nu}=\frac{h_{w}}{\sinh^{d} \rho} (d\tau^2+ d \rho^2-\frac{2}{d-2} \sinh^2 \rho d \Omega_{d-2}^2)
\end{equation}

We can now integrate it to obtain
\begin{equation}
\int_{S^1 \times H^{d-1}} \sqrt{g} \langle T_{\tau \tau} \rangle_W=- 2 \pi Vol(S^{d-2}) h_w \label{eq:EEhyp}
\end{equation}
where we  discarded  a divergent term going like $1/\epsilon $, where $\epsilon$ is a short
distance cutoff.
Thus, once we compute $h_w$ we can insert \nref{eq:EEhyp} into \nref{EntropyOne}.

\subsection{General features of the result}

Before we discuss more detailed results, let us discuss a general feature of the answer. If we consider
a Wilson loop in the fundamental representation in the large $N$ planar limit, then the first thing to
understand is the $N$ dependence for fixed $\lambda$.
In the literature, it is customary to include  a factor of $\frac{1}{N}$ in the definition of the
 Wilson loop. However, if we are considering the insertion of an actual physical quark, we should not
do this. In fact, we define the Wilson loop operator without this factor of $N$, $W = Tr[ e^{ i \int A + \cdots} ]$.
In this case the planar expectation value fo the Wilson loop will be $\langle W \rangle = N w(\lambda) $.
This follows from standard large $N$ counting arguments. When the theory is free, $\lambda =0$,
then the quark and antiquark have their indices correlated
 since the gauge invariant combination is
$|\Psi \rangle =  \frac{1}{\sqrt{N}} \sum_{i=1}^{N} |i\rangle | \bar{i} \rangle$. So this gives us a entanglement entropy $S_{\lambda=0}=\log N$.
This argument is a bit more subtle than it appears. The issue is that the unentangled states are
not gauge invariant!. In fact, this issue is related to the difficulty in defining entanglement entropy
for gauge fields (see \cite{Casini:2013rba} and references  therein). A reasonable way to define it is to separate the system into two and do not impose the
Gauss law on the entangling surface. Then the entanglement of color indices across this surface becomes
physical and contributes to the entanglement entropy, which for large $N$ gauge theories is just this $\log N$ factor.
This contribution is present in any planar large $N$ theory. At strong coupling it is related to a
a factor of $1/g_s$ in the partition function which is due to the fact that the Euclidean worldsheet has the
topology of a disk.
 It is just the contribution from the Einstein Hilbert term in the worldsheet
\begin{equation}
S_{g_s}=(n \partial_n-1) I_{EH}=(n\partial_n-1) \frac{\log g_s}{4 \pi} \int_{\Sigma_2} R= -\log g_s
\end{equation}
%We used that for a cone with conical excess $ 2\pi (n-1)$, the Euler characteristic is $\chi=1-n$ \cite{Fursaev:1995ef}.
  We used that the Euler characteristic of the disk is $\chi=1$.  This is essentially the Bekenstein-Hawking entropy on the string worlsheet. This contribution to the entropy is similar on spirit to that of \cite{Susskind:1994sm}.
Note that in any theory which contains strings, like large $N$ QCD, we will get
a $ - \log g_s \sim \log N $ contribution to the entanglement entropy. This is also true in two dimensional QCD.

For the particular case of  pure Chern-Simons, \cite{Kitaev:2005dm} computed the entanglement entropy
for this configuration. In that case $T_{\mu \nu} =0$ and the only contribution comes from the
expectation value of the Wilson loop , which is \cite{Witten:1988hf}
  $S_{W}=\log \dfrac{S^a_0}{S^0_0}=\log W$.
  $S^i_j$ is the modular $S$ matrix of the rational CFT.
At large $N$, this expression contains the $\log N$ we discussed above.

Once we take into account this factor of $N$, the rest depends on
the 't Hooft coupling $\lambda$.  At strong coupling we get a relatively large contribution
\cite{Jensen:2013ora} (see appendix \ref{sec:apb})
\be  \label{GravEntropy}
 S = { R^2 \over \alpha' (d-1) }
 \ee
which is of order $\sqrt{\lambda}$.
This is due to the classical area of the string, see figure \ref{stringhor}$(b)$.
More precisely,  the entropy is due to the change in the area when we change the temperature. At weak coupling
we also expect a similar effect. If we consider an external quark in a charged plasma, then we know
that it will produce a Yukawa style potential $ V = { e^2 \over r}e^{ - \mu r }$ where $\mu \sim
e T $ is the thermal mass. Then the self energy due to this potential is cutoff in the IR at $\mu$. This
introduces a temperature dependence in the free energy, and therefore a contribution to the entropy.

Then there can be  a term proportional to $\log R/l_s$ (or $\log \lambda $). This  a quantum correction on
the worldsheet and it comes from the quantum fields propagating on the string.
Since the worlsheet theory is a CFT with central charge $c = 12$ we expect the usual term
given by $ { c \over 6 } \log { R \over \epsilon } $ where $R$ is the IR cutoff and $\epsilon$ is a
short distance cutoff on the worldsheet, which could be $l_s = \sqrt{\alpha'}$. This gives a
candidate logarithmic term. However, we will see that this does not agree with the exact expressions we will derive
below.
 This means that
there must be other sources for logarithmic terms in the worldsheet computation that we have not identified. One source are
zero modes of the scalars on the internal manifold.
 Presumably a
careful analysis, similar to the one performed in  \cite{Sen:2012dw}, would produce a precise match for
these terms. We leave this problem to the future.  In fact, the worldsheet origin of these logarithmic terms has
never been properly explained for the circular Wilson loop.

\section{Free fields}

Here we are going to consider a free conformal scalar and vector field with line operators introduced. Our normalizations are
\begin{align}
{\cal L}_s=& \frac{1}{2} (\partial \phi)^2+\frac{1}{12} R \phi^2  \hspace*{0.75cm} W_s= e^{ e \int dt \phi(x(t))} \nonumber \\
{\cal L}_v=& \frac{1}{4} F^2 \hspace*{3cm}  W_v= e^{e i \int dx^{\mu} A_{\mu} }
\end{align}
For these free theories $h_w$  can be computed easily by considering for example a straight line. They are \cite{Kapustin:2005py} \footnote{Note that our definition for stress energy tensor differs in a sign so that our $h_w$ is positive.}
\begin{eqnarray}
h_{s}=\frac{e^2}{96 \pi^2} & ~~~~~~~~~~~~ h_v=\dfrac{e^2}{32 \pi^2}  \label{hValues}
\end{eqnarray}

\label{sec:n4check}
\vspace{0.5cm}

Now we are going to compute the entropies by calculating the contribution to the value of the Polyakov loop in the hyperboloid at temperature $\beta$.

\subsection{Entanglement entropy for a free  scalar field }
\vspace{0.5cm}

As  we showed in appendix \ref{sec:Aweak} (see also \cite{Cardy:2013nua}), the Green function at $\rho=0$ is ($\tau \sim \tau + \beta$):
\begin{equation}
G_{\beta}= (\frac{2 \pi}{\beta})^2\frac{1}{\cos {2 \pi\tau/\beta}-1}
\end{equation}
The scalar contribution to the partition function at any temperature is
\begin{equation} \label{betapartition}
\log W_{\beta}=\frac{e^2}{2} \int d\tau d \tau' G_{\beta}(\tau-\tau')=\pi e^2 \int_{0}^{2 \pi} \frac{1}{2(1-\cos \tau)}= 0
\end{equation}
where we substracted a divergent term.

Note that this is not compatible with the formulas \nref{OneTpoint} \nref{eq:EEhyp}.  In fact, the $\beta$ derivative of
 \nref{betapartition} is zero, and not given by   $h_s$ \nref{hValues}.
  This is because our definition of entanglement entropy that behaves nicely under conformal transformation
  is a bit different.
  The entropy computed from \nref{betapartition} does not
   take into account the regularization of the conical singularity that appears in the replica trick.
    The entropy that behaves more nicely under conformal transformations is the one defined by smoothing out the conical
    space that appears in the replica trick.
     The details are in appendix \ref{sec:wald} . The net result is the following. To the entropy computed from \nref{betapartition}
     we should add a term that comes from a Wald type term from the conformal coupling of the scalar to curvature. In the
      end we obtain
% REMOVELINE      This contact term is just the Wald entropy coming from the improvement term
\begin{equation}
S^s_W-\beta \partial_{\beta} \log W_{\beta}=\langle S_{wald} \rangle=-\frac{4 \pi}{12} \int dA \langle \phi^2 \rangle_{W}
\end{equation}

Where $A$ is the area of the tip of the cone. Now, $\langle \phi^2  \rangle_W= \dfrac{e^2}{16 \pi^2 }\sinh \rho^{-2}$ (see appendix \ref{sec:Aweak}), $A=4 \pi \sinh^2 \rho$ and the tip of the cone in hyperbolic coordinates is at $\rho=\infty$, so the $\sinh^2 \rho$ cancels and we have $\langle S_{wald} \rangle=-\dfrac{e^2}{12}$ .

So, indeed we get the expected result

\begin{equation} \label{WaldLike}
S_W^{s}=\log W+\int T_{\tau \tau}=-8 \pi^2 h_s=-\frac{e^2}{12}=\langle S_{wald} \rangle
\end{equation}
This is in agreement with  \nref{OneTpoint} \nref{eq:EEhyp}. Here $\log W =0$.

\subsection{Entanglement entropy for a free vector field }

As explained in the appendix \ref{sec:Aweak}, the contribution to the Wilson loop will only come from the temporal mode, which is simply a massless particle. When $\rho=0$, the zero temperature propagator
 is $G^{\infty}_{m^2=0}(t)=\dfrac{ K_1( t )}{4 \pi^2 t}$.

From here, the contribution to the Wilson loop is
 \begin{equation}  \label{Svector}
 \log W_{\beta}=-\frac{e^2}{2} \beta  \int_{- \infty}^{\infty} dt G_{\infty}(t)=\frac{e^2}{4} \frac{\beta}{2\pi}
 \end{equation}

Because the Polyakov loop is linear in $\beta$ it does not  contribute  to the entropy
\begin{equation}
S^v_W=0
\end{equation}
Indeed the circular Wilson loop expectation value is $\log W = { e^2 \over 4 }$ and this, together with
\nref{eq:EEhyp} \nref{hValues} gives \nref{Svector}.

\section{${\cal N}=4$ SYM results }

\label{sec:n4exact}

For ${\cal N}=4$ SYM we are going to consider the 1/2 BPS circular Wilson loop
\begin{equation}
W_{\circ}=\text{Tr}_{R} {\cal P} e^{i \oint \dot{x} \cdot A+ \oint n \cdot \Phi}
\end{equation}

Where $\dot{x}=d \tau$ and $n$ is a constant unit vector in the $S^5$.

Using localization \cite{Pestun:2007rz,Erickson:2000af,Drukker:2000rr} an exact expression for the loop was found
\begin{equation}
\langle W_{\circ} \rangle = L^{1}_{N-1}(-\frac{\lambda}{4 N}) e^{\frac{\lambda}{8 N}}
\end{equation}
The normalization of the stress energy tensor $h_w$ can be obtained by relating it \cite{Gomis:2008qa} with the normalization of a scalar operator of dimension $2$ which can be computed using localization \cite{Okuyama:2006jc} \footnote{This particular expression for the normalization of the operator is obtained when comparing the work of  \cite{Correa:2012at} and \cite{Fiol:2012sg}, and it is related with the Bremsstrahlung function of section \ref{sec:brem}.}

\begin{equation}
 h_w=\frac{1}{6 \pi^2} \lambda \partial_{\lambda} \log \langle W_{\circ} \rangle
\end{equation}

At weak coupling this just gives the sum of $h_w$ from the previous section and at strong coupling is the same that they obtain in \cite{Friess:2006fk} $T_{00} =\dfrac{\sqrt{\lambda}}{12 \pi^2 r^4}$.

\subsection{The entanglement entropy for a Wilson line insertion}

The exact result is then

\begin{equation}
 S_{W}=\log W_{\circ}-8 \pi^2 h_w=\left ( 1-\frac{4}{3} \lambda \partial_{\lambda} \right ) \log W_{\circ} \approx_{N \rightarrow \infty} \log \left ( \frac{2 N}{\sqrt{\lambda}} I_{1}(\sqrt{\lambda}) \right )-\frac{2}{3} \dfrac{\sqrt{\lambda} I_2(\sqrt{\lambda})}{I_1(\sqrt{\lambda})} \label{eq:n4ee}
\end{equation}

So
\begin{eqnarray} \label{eq:n4strong}
 S_{W}(\lambda \gg 1) &=& \log N+\frac{\sqrt{\lambda}}{3}-\frac{3}{4} \log \lambda+...  \\
 S_{W}({\lambda \ll 1})&=& \log N-\frac{\lambda}{24}+...
\end{eqnarray}

Recall that we are multiplying the usual Wilson loop by $N$. In the next section we are going to check that this is the expected result.

\begin{figure}[h!]
\begin{center}
\vspace{5mm}
\includegraphics[scale=0.7]{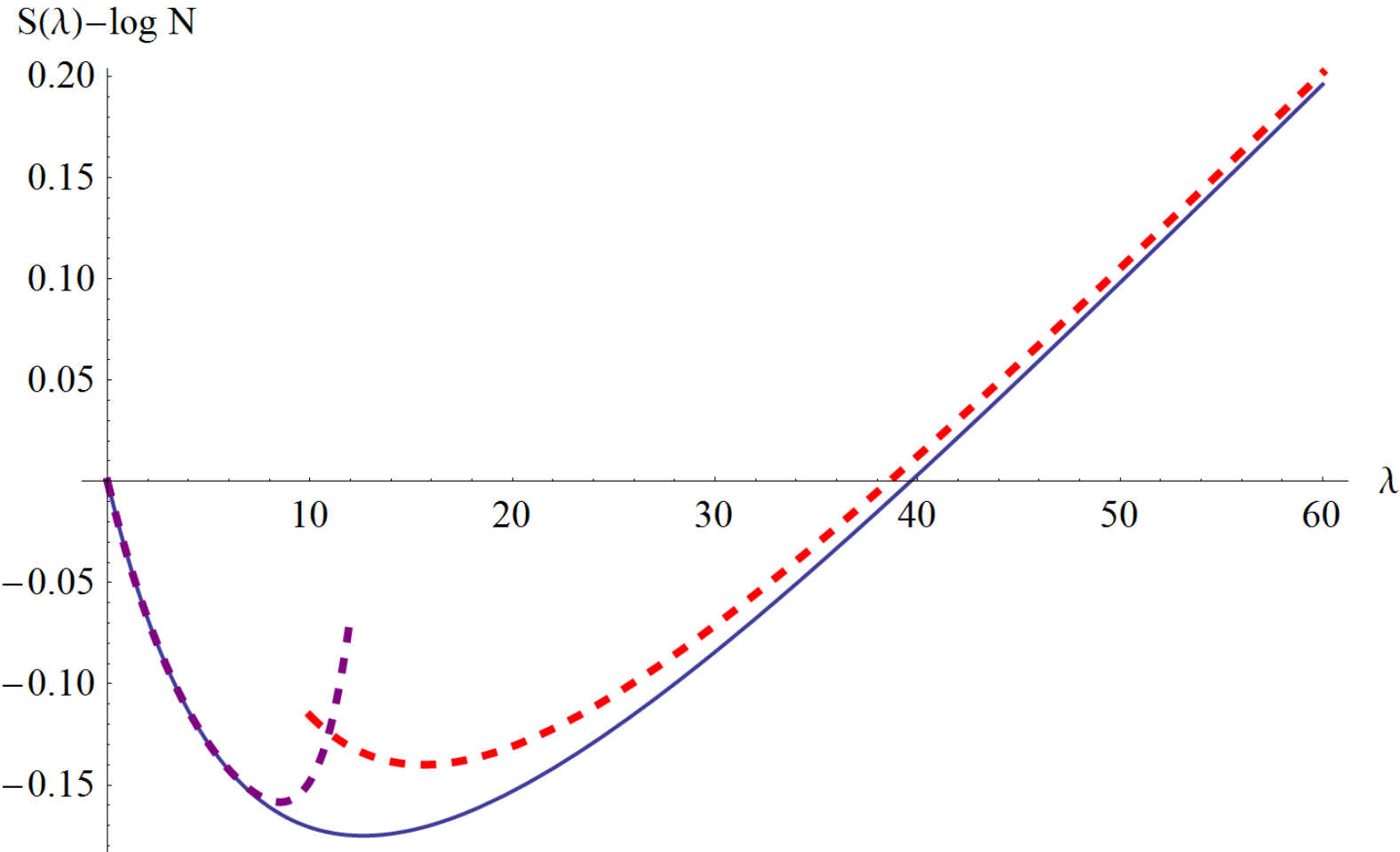}
\vspace{5mm}
\caption{
Entropy (solid blue) vs $\lambda$. We compare it with the weak coupling expansion (dashed purple) to order  $O(\lambda^{8})$ and with the strong coupling  result up to three loops (dashed red). Note that the weak coupling expansion has a finite
radius of convergence ($|\lambda| \sim 14.6$).  }
\label{figure:plot}
\end{center}
\end{figure}

 In figure \ref{figure:plot} we can see $S_W(\lambda)-\log N$ in the planar limit.

\subsection{Checks}

At weak coupling, the entropy is just the sum of the scalar and vector contribution

\begin{equation}
S_W^{{\cal N}=4}=(1-\frac{4}{3} \lambda \partial_{\lambda}) \log W=-\frac{\lambda}{24}=S_W^{s}+S_W^{v}
\end{equation}
We used that $e^2=\frac{\lambda}{2}$. This contribution comes from the Wald like term \nref{WaldLike}. Its negative
sign seems responsible for the initial decrease of the curve in figure \ref{figure:plot}.

% small typo
The leading term at strong coupling is the same as the entanglement entropy due to the classical string (see appendix \ref{sec:apb} for details).

 \subsection{Entanglement entropy for other branes}

\vspace{0.5cm}
{\bf D1 brane}
\vspace{0.5cm}

If we S dualize the exact result for the circular Wilson loop we obtain
\begin{equation}
\langle W \rangle_{1/g_{YM}}=e^{\frac{\pi ^2 N}{2 \lambda }} L_{N-1}^1\left(-\frac{\pi ^2 N}{\lambda }\right)
\end{equation}

 We can take the strong coupling limit, ie $g_s$ fixed. In this case we obtain
 \begin{equation}
 \langle W \rangle_{1/g_{YM}, \lambda \rightarrow \infty}=\frac{g_{YM} \lambda^{1/4} e^{\frac{2 \pi  \sqrt{\lambda }}{g_{YM}^2}}}{2 \pi ^2 }
 \end{equation}

    \vspace{0.5cm}
  {\bf Other representations}
  \vspace{0.5cm}

   We can compute the entanglement entropy for the $k^{th}$-symmmetric and  $k^{th}$-antisymmetric representations at strong coupling \cite{Drukker:2005kx,Yamaguchi:2006tq,Gomis:2006sb} for $k, N \to \infty$ and $k/N= $fixed. We find
\begin{eqnarray}
\log W_{A_k}=\frac{2 N}{3 \pi} \sqrt{\lambda} \sin \theta_k^3 \rightarrow S_{A_k}=\frac{2 N}{9 \pi} \sqrt{\lambda} \sin \theta_k^3 \\
\log W_{S_k}=2N (\bar{\kappa} \sqrt{\bar{\kappa}^2+1}+\sinh^{-1} \bar{\kappa}) \rightarrow S_{S_k}=2N (-\frac{2}{3} \bar{\kappa} \sqrt{\bar{\kappa}^2+1}+\sinh^{-1} \bar{\kappa})
\end{eqnarray}
Where $\bar{\kappa}=\frac{k \sqrt{\lambda}}{4 N}$ and $\sin \theta_k \cos \theta_k-\theta_k=\pi (\frac{k}{N}-1)$ .
For the first one we get the expected result since the D5 brane is wrapping the same $AdS_2$ surface as the fundamental string.
The second result seems less trivial. Checking it involves finding a D3 brane configuration in the background of a
 hyperbolic black hole with  $\beta \sim 2\pi $, but not just only the $\beta = 2 \pi $ case.

 \section{$3d$ ${\cal{N} }\ge 2 $ Chern-Simons - matter theories}

\begin{figure}[h!]
\begin{center}
\
\includegraphics[scale=0.6]{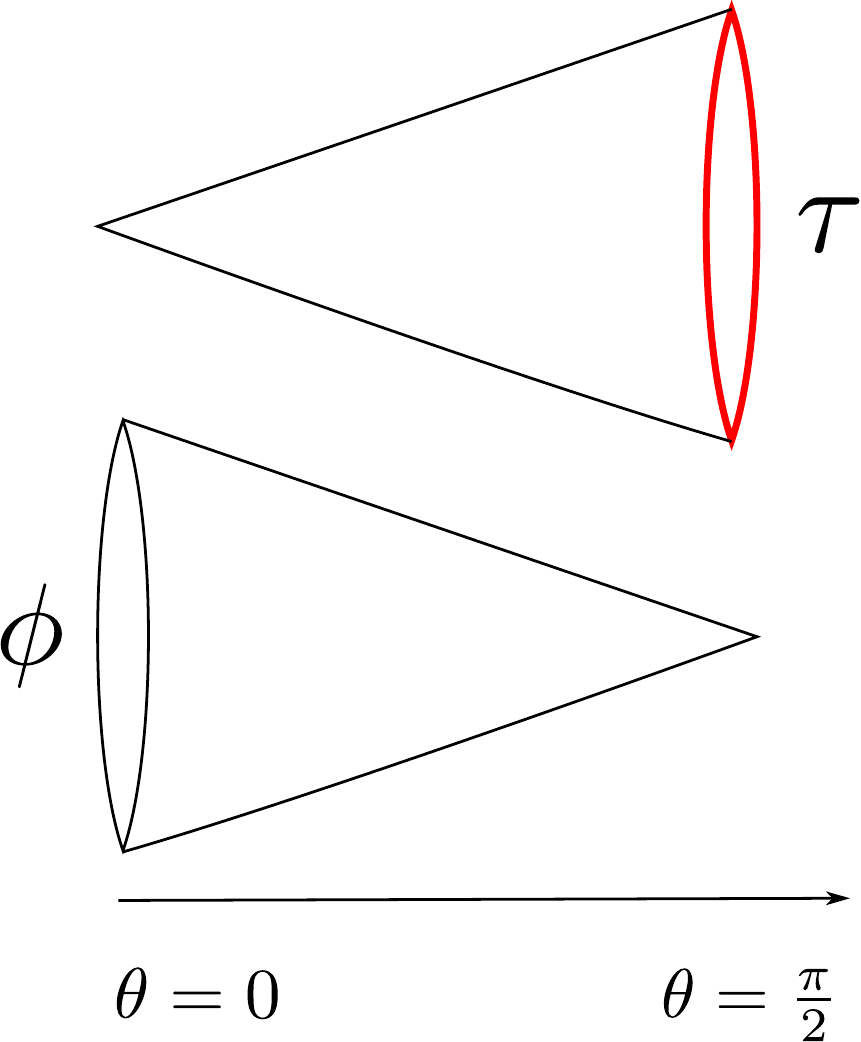}
\
\caption{The $\tau$ circle shrinks at $\theta=0$ while $\phi$ shrinks $\theta=\frac{\pi}{2}$. We are going to put a loop at $\theta=\frac{\pi}{2}$. The geometry is not singular, we just draw the cones to denote that the circles shrink.}
\label{fig:branch}
\end{center}
\end{figure}

 In \cite{Nishioka:2013haa}, they  defined a supersymmetric quantity similar to the R\'{e}nyi entropies in 3d. To do that, they considered the regularized (regularized in the sense of \cite{Fursaev:1995ef}) branched sphere (see figure \ref{fig:branch} and appendix \ref{sec:coord} for the relation with the previous metrics). A branched sphere is simply a sphere where one
 of the angles has its periodic identification changed from $2\pi$ to $2\pi n$.

  \begin{equation}
 ds^2=f_{\epsilon}(\theta)^2 d\theta^2+n^2 \sin \theta^2 d\tau^2+ \cos \theta^2 d\phi^2 \label{eq:regmet}
 \end{equation}

 Where $f_{\epsilon}(\theta \rightarrow 0)\rightarrow n,f_{\epsilon}(\theta>\epsilon)=1$ and $\tau \sim \tau+2\pi$. They applied the results of \cite{Closset:2012ru} to look for supersymmetric regularized spheres. They observed that this background can be supersymmetric by turning on a connection coupling to the $U(1)$ R symmetry
   \begin{equation}
   H=-\frac{i}{f_{\epsilon}} ; ~~~~~~ A^{(R)}=(\frac{n}{ f_{\epsilon}}-1) \frac{d\tau}{2}+(\frac{1}{ f_{\epsilon}}-1) \frac{d\phi}{2}
   \end{equation}
 where the killing spinor equation  $D_{\mu} \xi=-\frac{1}{2} H \gamma_{\mu} \xi$ is satisfied for two constant spinors with opposite R charges.
 For $n \not = 1$ the background is different that the original one but if $n\rightarrow 1,f_{\epsilon} \rightarrow 1$ we get the three sphere without any extra field turned on.

 After localizing a  Chern-Simons matter theory in this background they observed that the matrix model that they obtain is equivalent to the one of the squashed sphere \cite{Hama:2011ea} for $b=\sqrt{n}$. In \cite{Martelli:2011fu} it was also observed that in the localization computation nothing depends on the precise shape of $f$.

Now we would like to consider our Wilson loop in the $S^3$ and compute its entanglement entropy. The branched sphere is conformal to the hyperboloid, the loop we have been considering in these coordinates is extended along $\tau$ and sits at  $\theta=\frac{\pi}{2}$. Wilson loops in the $b$-deformed sphere were considered in  \cite{Kapustin:2013hpk,Tanaka:2012nr}.

   We can now compute the entanglement entropy for this Wilson loop \cite{Nishioka:2013haa}

   \begin{equation}
   S_W= \left. (1-n\partial_{n}) \log |W(S^3_{b=\sqrt{n}})| \right|_{n=1}
   \end{equation}
Here we have taken the absolute value to remove an unexpected phase.
The phase is  unexpected because the computation of a Wilson loop expectation value
 has an interpretation as the norm of a state. This is
the state produced by Euclidean evolution on a half sphere with the insertion of a half circular  Wilson line.
The entanglement entropy  computation  also has a similar interpretation.
A similar phase is also present in the computation of the sphere partition function. It was shown in
\cite{Closset:2012vg} that it arises from a local counterterm because the regularization that is implicitly used in
performing the localization procedure does not respect the unitarity of the theory.
Here we expect a similar interpretation. For this reason we are always
going to ignore such phases. For the pure Chern Simons theory this goes under the name of ``framing ambiguity''.

 The supersymmetric Wilson loop  that we are considering on  the squashed sphere is
 \begin{equation}
W(S^3_b)={\text Tr} {\cal P} e^{\oint i A+\sigma}
\end{equation}
Where $\sigma$ is the scalar partner in the vector multiplet. These theories in the squashed sphere localize to a matrix model \cite{Hama:2011ea}, so the expectation value of the Wilson loop is just the expectation value in a matrix model: $\langle W(S^3_b) \rangle= \langle e^{2 \pi b \sigma}\rangle_{S^3_b,MM}$.

 An analytic expression seems complicated to find. However, because we are only interested in the first derivative at $b=1$, we can make the problem easier. This is because  the Chern-Simons and matter terms in the matrix model are symmetric under $b \rightarrow b^{-1}$ and thus $b$ even to first order.
 This means that the only contribution will come from the localization of the Wilson loop, ie we can set $b=1$ for the other factors and compute the expectation value of the winding Wilson loop in $S^3$. So we only have to compute the expected value of the winding Wilson loop, do the analytical continuation and take the derivative:
   $\partial_b  W(S_b^3)
| _{b=1} =\partial_m \langle  e^{2 \pi m \sigma} \rangle_{S^3,MM}|_{m=1}$ . We write $m$ so it is clear that it is a loop wound $m$ times in a great circle of the $S^3$.

   Before exploring the case of ABJM, we will analyze its relation with $h_w$.

   \vspace{0.5cm}
  { \bf Relation with $h_w$}
\vspace{0.5cm}

 If we now expand the expectation value of the loop to linear order in $n-1$ , we get \cite{Closset:2012ru}:
 \begin{equation}
 -\partial_n \log W=\int_{S^3} -\frac{1}{2}\langle T_{\mu \nu} \rangle_{W,S^3 } \partial_n g^{\mu \nu}+\langle j_{\mu}^R\rangle_{W,S^3 }  \partial_n  A^{\mu}+\langle J^{Z}\rangle_{W,S^3 } \partial_n  H=-\int_{S^3} \frac{1}{2} \langle T_{\mu \nu} \rangle_{W,S^3 } \partial_n  g^{\mu \nu}
 \end{equation}

The term $\langle j_{\mu}^R\rangle_{W,S^3 }$ drops out
because the scalar $\sigma$ does not carry $R$  charge.
 And the other term cancels because for the regularized sphere $\partial_n H \sim \partial_n f_{\epsilon}$ is localized in a small region near $\theta=0$ which ends up having measure zero as $\epsilon \rightarrow 0$. Also the operator $J^Z$ is zero for 
  a CFT. 
  Note that the overall sign is positive because $\partial_n g^{\tau \tau}=-2$.

Because of the relation of this derivative with the wound loop, we can compute $h_w$ from it\footnote{
Note that since we are smoothing out the cone, we are automatically computing the entropy that includes possible
Wald like contributions. }
\begin{equation}
4 \pi^2 h_w= \left. \frac{1}{2} \partial_m \log |\langle W_m \rangle_{S^3}| \right|_{m=1}
\end{equation}

\subsection{ABJM}
\label{sec:ABJM}

For ABJM two kinds of circular loops are known \cite{Drukker:2008zx,Drukker:2008zx2,Drukker:2008zx3,Drukker:2008zx4}, they are 1/6 and 1/2 BPS respectively\footnote{See \cite{Cardinali:2012ru} for a discussion on more general contours.}. Here we discuss the former, in appendix \ref{sec:ABJMw} there is a discussion about the 1/2 BPS loop. If we group the scalars into $C$, which transforms in the fundamental of $SU(4)_R$, this Wilson loop can be written as
\begin{equation}
W^{1/6}={\text Tr} {\cal P} e^{\oint i A+\frac{2\pi}{k} M_{I J} C^{I} \bar{C}^{J}}
\end{equation}

Where  $M=\text{diag}( 1,1,-1,-1 )$ and the contour is the great circle of $S^3$. This is the same loop that we considered in the previous section.  When we integrate out $\sigma$ we get $\sigma=\frac{2\pi}{k} M_{I J} C^{I} \bar{C}^{J}$ (see (4.11) of \cite{Benna:2008zy} for more details).
This means that the loops we are considering are  $1/6$ BPS loops.
From the string theory perspective, the string configuration which preserves 1/6 SUSY  has Neumann boundary
 conditions along a $CP^1 \subset CP^3$ \cite{Drukker:2008zx}. Thus,  there are two zero modes.

The multiply wound Wilson loop in ABJM was analyzed in \cite{Klemm:2012ii}, we will denote a loop that winds $m$ times the great  circle $W^{1/6}_m$.

They have an expression which can be expanded at strong coupling \cite{Klemm:2012ii}

\begin{equation}
\langle W_m^{1/6} \rangle=\frac{2 \pi i^{m+1}}{g_s} e^{\pi m \sqrt{2 \lambda}} \left (\frac{\sqrt{\lambda}}{2 \sqrt{2} \pi m}-(\frac{H_m}{4 \pi^2 m}+\frac{i}{8 \pi m}+\frac{1}{96} )+(\frac{i}{192}+\frac{\pi m}{4608}+\frac{H_{m-1}}{96 \pi} ) \frac{1}{2 \sqrt{\lambda}}+O(\lambda^{-1}) \right)
\end{equation}

Where $ \lambda=\frac{N}{k}=\frac{R^4}{2 \pi^2 \alpha'^2}$ and $H_m$ is the $m$-th harmonic number.

Now, using the entropy formula $S_W=(1-\frac{1}{2} m \partial_m)|\log W^{1/6}_m|_{m=1}$, we obtain
\begin{equation}
S^{1/6}_W=-\log g_s+ \frac{R^2}{2 \alpha'} + \log \frac{R^2}{ \alpha'}+O(1)
\end{equation}
We expressed it in term of $R,\alpha',g_s \propto \frac{\lambda}{N}$ to compare with the previous worldhseet
calculations, ie $\sqrt{2 \pi^2 \lambda}=\frac{R^2}{\alpha'}+O(\frac{1}{\sqrt{\lambda}})$.
The leading contribution agrees with the gravity result \nref{GravEntropy}.

 The weak coupling expansion can be easily done (in the planar limit) using the matrix model of \cite{Marino:2009jd,Drukker:2010nc}

 \begin{equation} \label{ABJMMM}
 \langle W_m^{1/6} \rangle= \frac{1}{2 \pi^2 i \lambda} \int_{-a}^{a} dx e^{m x} \arctan \sqrt{\frac{\alpha-2 \cosh x}{\beta+2 \cosh x}}
 \end{equation}
Where
\begin{align} \label{kappa}
\lambda & = \frac{N}{k}=  { \kappa \over 8 \pi } ~ {}_3F_2( { 1 \over 2 } , {1 \over 2} ,{1 \over 2} ; 1,{ 3 \over 2} ; - { \kappa^2 \over 16 } )  \nonumber \\
e^a & = { 1 \over 2} \left( 2 + i \kappa + \sqrt{ \kappa ( 4 i - \kappa ) } \right) \nonumber \\ \alpha & =  2 + i \kappa \hspace{1cm}  \beta = 2 -  i \kappa
\end{align}

 If we expand this integral for small $\lambda$  we obtain for the entropy
\begin{equation}
 S_W^{1/6}=\log N-\frac{ \pi ^2 }{6}\lambda ^2+\frac{37\pi ^4}{90}  \lambda ^4-\frac{2887\pi ^6 }{4536} \lambda ^6+\frac{288937\pi ^8 }{302400}  \lambda ^8-\frac{5380427\pi ^{10} }{3742200}  \lambda ^{10}+O(\lambda^{12})
\end{equation}

\begin{figure}[h!]
\begin{center}
\vspace{5mm}
\includegraphics[scale=0.7]{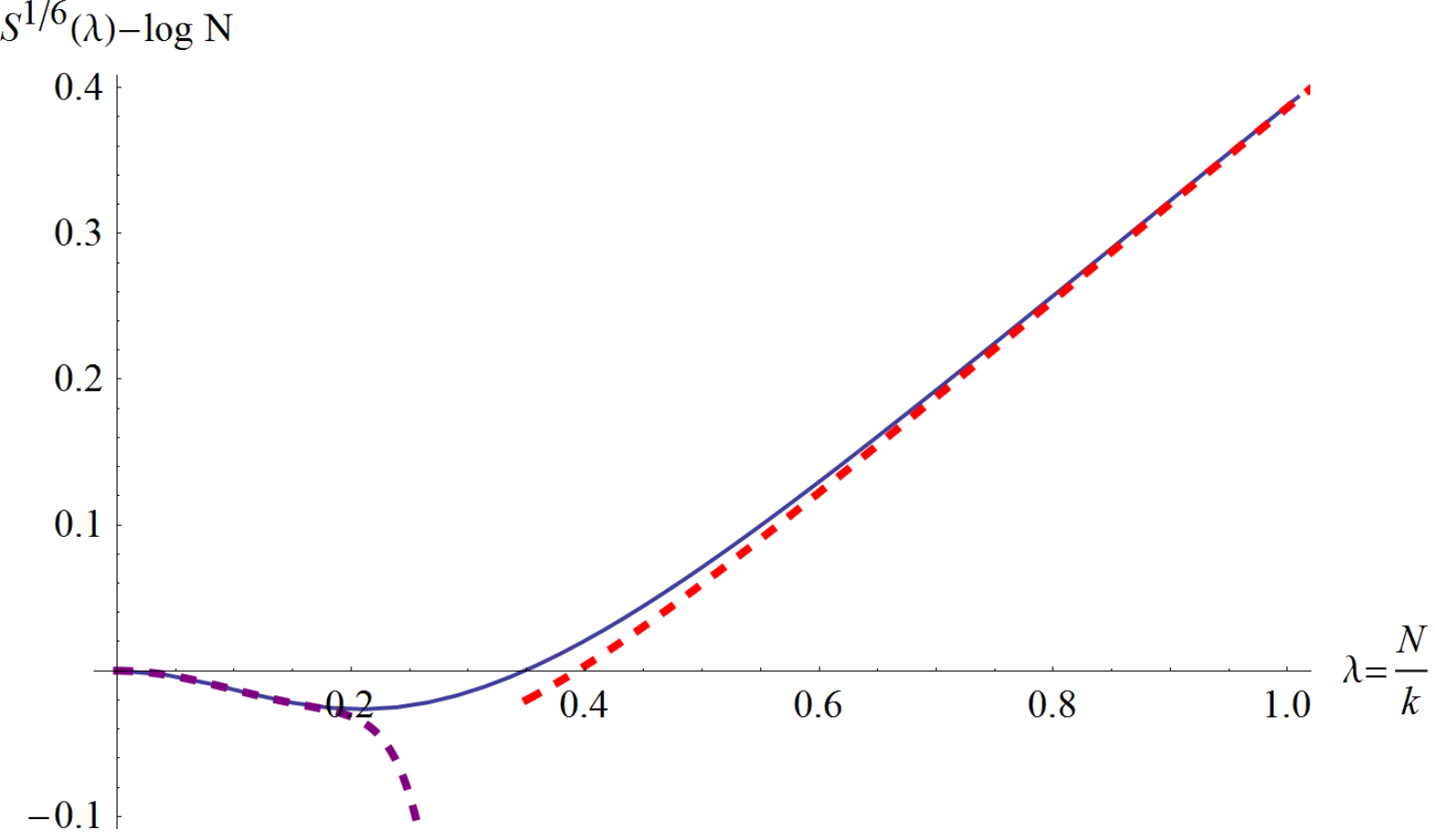}
\vspace{5mm}
\caption{
Entropy (solid blue) of the 1/6  BPS loop in the planar limit. Weak coupling (dashed purple) up to $O(\lambda^{10})$ and strong coupling (red purple) at three loops. }
\label{fig:abjmplotS}
\end{center}
\end{figure}

In figure \ref{fig:abjmplotS} we can see how the exact planar result  interpolates between weak and strong coupling.

\section{Relation between conformal stress energy tensor and  the Bremsstrahlung function}

\label{sec:brem}

The energy radiated by an accelerated quark has the form $ E  = 2 \pi B \int dt  \dot v^2 $ for small velocities.
The general formula has the same form but we replace $\dot v$ by the proper acceleration. Here $B$ is a function of
the coupling constant.
Since we are considering a configuration that involves an accelerating quark-antiquark pair it seems that we can
relate the radiated energy to a flux of the stress tensor. Thus we expect a relation between the coefficient $h_w$ that sets
the value of the stress tensor and the function $B$.
In this section we discuss the relation between the two. It turns out that there is no universal relation since we can
compute these two things independently for various theories, such as a free Maxwell field, or a free scalar field, or
weakly coupled ${\cal N} =4$ super Yang Mills and we get a different ratio for $h_w/B$ in each of these cases\footnote{  For the free fields the Bremsstrahlung functions are $B^s=\dfrac{e^2}{24 \pi^2}; B^v=\dfrac{e^2}{12 \pi^2}$ (see \cite{Athanasiou:2010pv} for example). The corresponding $h_w$ values are in \nref{hValues}.}.
We think that this is related to the problem of separating the radiation component from the self energy part of the field.
If we had zero acceleration in the far past and far future this would not be an issue. However, it is an issue when
we have constant acceleration and there has been quite a bit of discussion in the literature on this point \cite{Boulware:1979qj}.

\begin{figure}[h!]
\begin{center}
\vspace{5mm}
\includegraphics[scale=1.3]{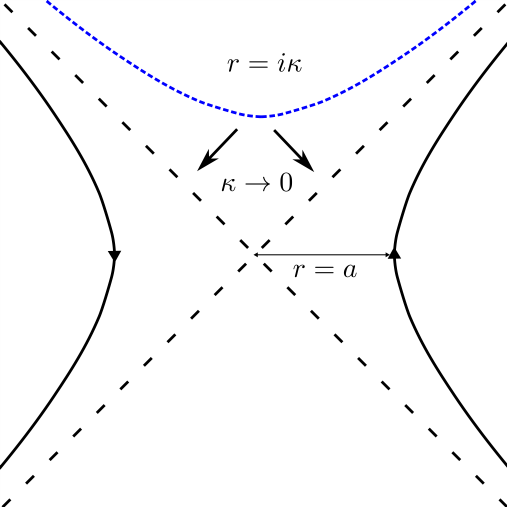}
\vspace{5mm}
\caption{
We  are  going to integrate the stress energy tensor over the surface of constant $\kappa$ and then send it to zero.}
\label{Brems}
\end{center}
\end{figure}

We consider the metric
\be \label{metrpol}
ds^2 = - r^2 d\tilde \tau^2 + dr^2 + dy^2 + y^2 d\Omega_{d-3}^2
\ee
We set the quark at $r=a$ and $y=0$. Here the ordinary Minkowski time $t$ is given by $ t = r \sinh \tilde \tau $.
By analytically continuing $r \to i \kappa$ we can go into the Milne region and compute the flux of energy through the
surface at a fixed $\kappa$, see figure \ref{Brems}. The Killing vector associated to Minkowski energy is
$ \zeta^\mu \partial_\mu = \partial_t$. Then we need to integrate $\int_{\Sigma_{d-1} } * j $, where $j_\mu = T_{\mu \nu} \zeta^\nu $. Expressing the Killing vector in the coordinates \nref{metrpol} we find
\be \label{enrad}
E = \int_{ \kappa = {\rm const}} * j = \left.
\int^{\tilde \tau_{\rm max} } d\tilde \tau \kappa \cosh \tilde \tau \int dy y^{d-3} d\Omega_{d-3} T^{rr}
\right|_{\kappa \to 0 }
\ee
where the cutoff in the proper time integral is chosen so that $t_{\rm max} = \kappa \sinh \tilde \tau_{\rm max }
\sim \kappa \cosh \tilde \tau_{\rm max} $ (where $\tilde \tau_{\rm max} \gg 1 $).
 Thus the first integral gives a factor of $t_{max}$. Note that we are integrating
between roughly $\tilde \tau \sim 0$ and the large value $\tau_{max}$. The integral over negative values is expected to
correspond to the radiation emitted by the other particle. The expression for $T^{rr}$ is
\be \label{trrdef}
T^{rr} = h_w \left( { 1 \over \tilde r^{d} }  + {\rm not~ important} \right) ~,~~~~~~~\tilde r  = { \sqrt{ (r^2 + y^2 -a^2)^2 + 4 a^2 y^2 }
\over 2 a }
\ee
where we did not specify the terms that vanish in the $\kappa \to 0$ (or $r \to 0$) limit. We can easily now set $\kappa =0$ (or $r=0$) and
integrate over $y$ to obtain
\be \label{BNaive}
E = {  t_{max} \dot v^2  }  (2 \pi )    { d \pi^{\frac{d-3}{2}} \over 2 \Gamma( { d +1  \over 2 }   ) } h_w  ~,~~~~~~~\dot v^2 = { 1 \over a^2 }
\ee

In general this gives us a different answer than the function $B$ computed directly\footnote{For a maxwell field in $d=4$ it gives
us the same answer.}. We think that the reason for the disagreement is the improper separation between the radiated energy and
the self energy.

In supersymmetric theories, such as ${\cal N}=4$ super Yang Mills, or Chern Simons matter theories,
 we can subtract the
self energy in the following way. In those theories we have a dimension $\Delta = d-2$ scalar operator that gets an
expectation value in the presence of a Wilson loop. Then we define a new conserved (but not traceless)
stress tensor, $\tilde T$ by adding a total derivative
  \begin{equation}
\tilde{T}_{\mu\nu}=T_{\mu\nu}+\alpha (g_{\mu \nu} \nabla^2-\nabla_{\mu} \nabla_{\nu}) O_{d-2}
\end{equation}
where $T_{\mu\nu}$ is the standard traceless stress tensor.
Here $\alpha $ is defined so that the stress tensor $\tilde T$ has no $ { 1/\tilde r^{d}}$ singularity at short distances
from the loop. Setting the normalization of $O_{d-2}$ to $\langle O_{d-2} \rangle = { h_w \over \tilde r^{d-2} }$, with
$\tilde r$ as in \nref{trrdef}, we find $\alpha = 1/(d-2)^2$.
We can now do the same integral as in
\nref{enrad}, but using $\tilde T^{rr}$ to find
\begin{equation}
E = \left.
\int^{\tilde \tau_{\rm max} } d\tilde \tau \kappa \cosh \tilde \tau \int dy y^{d-3} d\Omega_{d-3} \tilde T^{rr}
\right|_{\kappa \to 0 } \nonumber  =\frac{2 \pi t_{max}}{a^2} \frac{\pi^{\frac{d-3}{2}}}{2 \Gamma(\frac{d+1 }{2} )} (d + (d-2) \alpha)h_w
\ee

Therefore
\begin{equation} \label{Bfunction}
\tilde{B}=\frac{4 {\pi}^{d+1\over 2}}{ \Gamma(\frac{d-1}{2})} \frac{d-1}{d-2} \frac{h_w}{4 \pi^2}
\end{equation}
The tilde just means that we have defined this function simply in terms of the radiated energy using the above procedure.
We conjecture that $\tilde B = B$ for these theories.  In fact, we can check that at strong coupling we can compute
both $h_w$ and $B$ independently and find that the answer is in agreement with \nref{Bfunction} in any dimension. This can be done as follows. From the expression for the entropy $S={ R^2 \over \alpha' ( d-1)}$
we can compute $h_w$ by using the circular Wilson loop expectation value $\log \langle W \rangle = {   R^2 \over \alpha' }$ and \nref{EntropyOne}, \nref{eq:EEhyp}. The computation of $B$ using the classical
string worldsheet  the same in all
dimensions and gives $B = { R^2 \over 4 \pi^2 \alpha' }$, in agreement with \nref{Bfunction}.
For ${\cal N}=4$ SYM this relation \nref{Bfunction} was noticed in \cite{Fiol:2012sg}.
In \cite{Correa:2012at}, $B$ was computed using other relations among
 supersymmetric Wilson observables. Note that \nref{Bfunction} is bigger than the naively
obtained expression from \nref{BNaive}.

There is probably a more rigorous logic for deriving this result. At this point it is just a reasonable
conjecture, backed up by a qualitative argument. The idea is that we need to subtract the self energy
contribution in some intrinsic way. Subtracting it using the expectation value of the operator
$O_{d-2}$ is a reasonably intrinsic way to do it. A precise derivation could perhaps involve the
appearance of $\tilde T$ in the right hand side of supersymmetry anticommutation relations. We leave
such a derivation to the future.

It is interesting to look at the problem of computing the radiation at strong coupling.
The string  worldsheet in AdS is described by $  - t^2 + x^2 + z^2 = a^2$ in Poincare coordinates.
We can compute the spacetime energy at a given time
by integrating the corresponding current on the worldsheet.
We get the right expression for the radiation by integrating this current over the portion of the worlsheet
that is within the  region  $t^2-x^2 \geq 0$ and any $z$.
 If we consider Rindler-AdS space, this is the region
of the worldsheet that lies behind the Rindler AdS horizon. This was done in
\cite{Mikhailov:2003er,Xiao:2008nr}. See also \cite{Chernicoff:2011vn}.

  At strong coupling, one can also compute the power radiated from the spacetime perspective (by integrating the stress energy tensor). \cite{Hatta:2011gh} did it for an arbitrary trajectory. They observed that the power radiated had two pieces, one which didn't contribute when the quark began and ended at rest and they claimed that this term shouldn't be identified with radiation. For the case of uniformly accelerated motion, this term is exactly the difference between our naive and improved radiation. This is an explicit example of the fact that it is non trivial to separate the self-energy from the radiation unless the quark ends with constant velocity.
 See also \cite{Agon:2014rda} for a recent discussion\footnote{We thank Alberto Guijosa for discussions on these matters.}.

\subsection{Bremsstrahlung function in ABJM}

For ABJM,  we have $O_{1}=\frac{2 \pi}{k} M_{I J} C^{I} \bar{C}^{J}$ which sits in the stress tensor multiplet. Now, using the result of the previous section we get $B=2 h_w$.

Using the exact result from section \ref{sec:ABJM} for the 1/6 BPS Wilson loop we have that the Bremsstrahlung function in ABJM in the planar limit is then

\begin{align}
B= & \left. \frac{1}{4 \pi^2} \partial_m \log |W_m| \right|_{m=1}
\end{align}
Where $W_m$ is the Wilson loop wrapping the great circle of the $S^3$ wound $m$ times. This was computed exactly and, as we saw in section \ref{sec:ABJM} is just an integral. We can expand at weak and strong coupling
 \begin{align}
B^{1/6}=_{\lambda \ll 1} & \frac{\lambda ^2}{2}-\frac{\pi ^2 \lambda ^4}{2}+\frac{47 \pi ^4 \lambda ^6}{72}-\frac{17 \pi ^6 \lambda ^8}{18}+\frac{73667 \pi ^8 \lambda ^{10}}{50400}  \nonumber \\
B^{1/6}=_{\lambda \gg 1}  & \frac{\sqrt{2 \pi^2 \lambda }}{4 \pi ^2}-\frac{1}{4 \pi ^2}+\left(\frac{1}{4 \pi ^2}-\frac{5}{96}\right) \sqrt{\frac{1}{2 \pi^2 \lambda }}
\end{align}

In figure \ref{fig:abjmplotB1} we plotted the function.

\begin{figure}[h!]
\begin{center}
\vspace{5mm}
\includegraphics[scale=0.7]{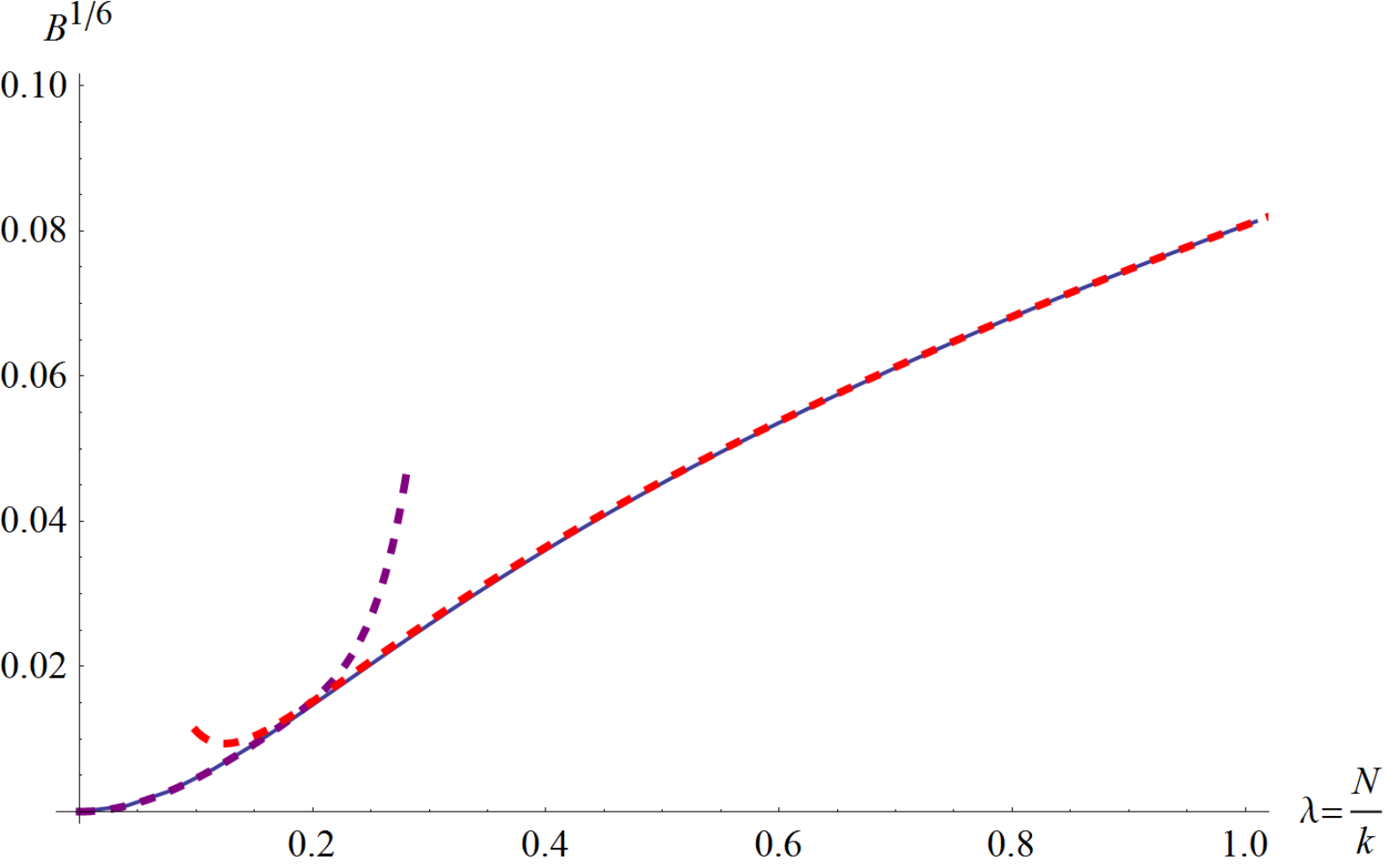}
\vspace{5mm}
\caption{
Bremsstrahlung function (solid blue) for ABJM, 1/6  BPS loop in the planar limit. Weak coupling (dashed purple) up to $O(\lambda^{10})$ and strong coupling (dashed red) at three loops. The strong coupling expansion is extremely good. }
\label{fig:abjmplotB1}
\end{center}
\end{figure}

{\bf Relation with cusp anomalous dimension}

 As shown in \cite{Correa:2012at}, we can relate the Bremsstrahlung function and the cusp anomalous dimension in conformal field theories. The displacement operator will also have dimension $2$ in $3d$, so $\Gamma_{cusp}\sim -\phi^2 B$ should still be true in $3d$.

At this moment there is no cusp anomalous dimension computed at weak or strong coupling for the 1/6 BPS Wilson loop. As we said before, the leading order at strong coupling is the same as the energy radiated by the string.  In the appendix \ref{sec:ABJMw} there are some comments about a possible $B$ function for the 1/2 BPS loop.

\section{Conclusions}

In this note we have used the localization techniques of \cite{Pestun:2007rz,Kapustin:2009kz,Nishioka:2013haa,Marino:2009jd,Klemm:2012ii} to compute the
entanglement entropy of a spherical region containing a Wilson line insertion in
${\cal N}=4$ SYM and ABJM.
We have also given a candidate expression for the
Bremsstrahlung function in supersymmmetric theories. The derivation of this function is partly conjectural
but it seems quite likely to be correct.
 This prescription reproduces
 the known result  in \cite{Correa:2012at,Fiol:2012sg} for ${\cal N}=4$ SYM and gives a new proposal for
 BPS Wilson loops in general ${\cal N} =2$ theories, including the ABJM theory. For the case of
the ABJM theory we obtained a result for the 1/6 BPS Wilson loop. For the 1/2 BPS Wilson loop we
did not obtain a reasonable answer, as discussed in appendix \ref{sec:ABJMw} .

In summary, we have related $S_W-\log \langle  W\rangle ,h_w$ and $B$ in these theories.

We would like to end by mentioning some open problems.

When these formulas are expanded at strong coupling, they contain terms that can be interpreted as
the entanglement entropy of the string worldsheet across the horizon.  The leading term comes from
the classical action of the string. But the subleading correction should contain some information about
the quantum entanglement on the string worldsheet.  This should be contained in the $\log \lambda$
terms. However, we could not properly match the coefficients. It would be nice to match them.

As mentioned in the introduction, it would be nice if the cusp anomalous dimension could also be
 computed using integrability techniques. In this way one could obtain the function $h(\lambda)$. An  expression for $h(\lambda)$ was  recently proposed in \cite{Gromov:2014eha}. It would be also be good to obtain matching for $B$ at one loop at strong coupling, presumably by using the proper regularization for the fluctuations of the string.  Another interesting thing would be a more formal derivation (maybe using the SUSY algebra) of the prescription we used  to compute the radiation in the presence of conformal scalar.

{\bf Note:} While this note was being prepared, \cite{Marmiroli:2013nza} proposed a Bremsstrahlung function for ABJM. His expressions for the 1/6 BPS Wilson loop differ from ours. The expressions in
\cite{Marmiroli:2013nza} are complex, but even the real part is  different. For the 1/2 BPS Wilson loops he
obtained an expression given by even powers of $\lambda$. These differ from explicit computations
at weak coupling, as discussed in appendix \ref{sec:ABJMw}.

\section{Acknowledgements}

We would like to thank D. Correa, M. Leoni, T. Nishioka,
J. Preskill, H. Verlinde, B. Willett and I. Yaakov for discussions. JM was supported in part by U.S. Department
of Energy grant DE-SC0009988. AL acknowledges support from ``Fundacion La Caixa".

\newpage
\appendix
\section{Coordinate systems}
\label{sec:coord}

{\hspace{0.5cm} \bf Hyperboloid}

Metric: $ds_{S^1 \times H^{d-1}}^2=d \tau^2+d\rho^2+\sinh \rho^2 (d \phi^2+ \sin^2 \phi d\Omega_{d-3}^2)$

Wilson loop: along $\tau$ at $\rho=0 $

Expected value primary scalar operator $\langle {\cal O} \rangle_W =\dfrac{h_{\cal O}}{\sinh ^{2 \Delta}\rho}$

{\bf Double polars}

Relation with hyperboloid: $ds^2=r^2 ds_{S^1 \times H^{d-1}}^2$ , $\cosh \rho = \frac{a^2+y^2+r^2}{2 r a} , \text{cotan} \theta = \frac{a^2-y^2-r^2}{2 y a}$

Metric: $ ds^2=r^2 d \tau^2+dr^2+dy^2+y^2 d\Omega_{d-3}^2$

Wilson loop:  along $\tau$  at $r=a $

Expected value primary scalar operator $\langle {\cal O} \rangle_W =\dfrac{h_{\cal O}}{\tilde{r}^{2 \Delta}}$,  $\tilde{r}=\dfrac{\sqrt{(r^2+y^2-a^2)^2+4 a^2 y^2}}{2 a}$

{\bf Sphere}

Relation with hyperboloid:  $ds^2=\sin^2 \theta ds_{S^1 \times H^{d-1}}^2, \text{cotan} \theta = \sinh \rho$

Metric: $ds^2=\sin^2 \theta d \tau^2+d \theta^2+\cos^2 \theta  (d \phi^2+ \sin^2 \phi d\Omega_{d-3}^2)$

Wilson loop: along $\tau$  at $\theta=\frac{\pi}{2} $

Expected value primary scalar operator $\langle {\cal O} \rangle_W =\dfrac{h_{\cal O}}{\cos^{2 \Delta} \theta}$

{\bf Straight line}

Relation with hyperboloid:  $ds^2=(\cosh \rho + \cos \tau )^{-2} ds_{S^1 \times H^{d-1}}^2,$\\ \hspace*{5mm} $  t=\dfrac{\sinh \tau}{\cosh \rho + \cos \tau},r=\dfrac{\sinh \rho}{\cosh \rho + \cos \tau}$

Metric: $ds^2=dt^2+dr^2+r^2 (d \phi^2+ \sin^2 \phi d\Omega_{d-3}^2)$

Wilson loop: along $t$  at $r=0 $

Expected value primary scalar operator $\langle {\cal O} \rangle_W =\dfrac{h_{\cal O}}{r^{2 \Delta}}$

{\bf Stress tensor}

For all the systems where the conformal invariant distance $l$  ($\langle {\cal O} \rangle \sim l^{-2 \Delta}$) only depends on one coordinate $X$, then the stress energy tensor is diagonal and is given by:
\begin{equation}
\langle {T^{\mu}}_{\nu} \rangle_W  dx^{\nu}\partial_{\mu} =\frac{h_w}{l^{2 d}} ( dx^{\tau}\partial_{\tau}+ dx^{X} \partial_{X}-\frac{2}{d-2} \ dx^{\phi_i} \partial_{\phi_i})
\end{equation}
Where are the angles $\phi_i$ parametrize the $d-2$ sphere.

For the circular loop in double polars in $4d$, the expression can be found in \cite{Gomis:2008qa}
(note that our $h_w$ has an extra minus sign, so is positive definite):
\begin{align}
 T_{rr}& =h_w \left ( \frac{1}{\tilde{r}^4}-\frac{2 r^2 y^2}{a^2 \tilde{r}^6} \right ), & T_{\tau \tau}& =h_w \frac{r^2}{\tilde{r}^4} \nonumber \\
   T_{yy}& =h_w \left ( \frac{1}{\tilde{r}^4}-\frac{(a^2+y^2-r^2)^2}{2 a^2 \tilde{r}^6} \right ), & T_{\phi \phi} & =-h_w \frac{y^2}{\tilde{r}^4} \nonumber \\ T_{r y}& =-h_w \frac{ r y (a^2+y^2-r^2)}{a^2 \tilde{r}^6} & &
\end{align}

\section{Strong coupling calculation of entropy}
\label{sec:apb}
First consider general metrics
\begin{equation}
ds^2 = - h dt^2  + dr^2/h + \cdots
\end{equation}
the temperature is
\begin{equation}
\beta = 2 \pi { 2 \over h'(r_0) } ~,~~~~~~h(r_0)=0
\end{equation}
Now we consider a string in the background of this black hole. Its action is given by
\begin{equation}
 \log Z  = - \frac{1}{2 \pi \alpha'} \beta \int_{r_0}^\infty   dr =  \frac{1}{2 \pi \alpha'}  \beta r_0  + \beta { ( \rm  divergent )}
\end{equation}
 where the divergent part does not contribute to the entropy.
 We can now compute the entropy from this by taking
\begin{equation}
 S = (1- \beta \partial_\beta ) \log Z = - \frac{1}{2 \pi \alpha'} \beta   \beta { \partial r_0 \over \partial \beta} =
   \frac{1}{2 \pi \alpha'} \beta  { h'(r_0) \over (h'(r_0))' } = \frac{2}{\alpha'} { 1 \over ( h'(r_0))'}
\end{equation}
  Notice that $h'$ means the derivative with respect to $r$. While the second prime, means derivative
  with respect to $r_0$
\begin{equation}
  ( h'(r_0))' \equiv   { d \over d r_0 }  \left( \left. { d h \over d r }\right|_{r=r_0} \right)
 \end{equation}

{\bf Planar black brane}

We get $S = \frac{ 2 R^2}{\alpha' d } $.

\vspace{5mm}
{\bf Hyperbolic black brane}

In this case $h = r^2 - 1 - \mu/r^{d-2} $.
In our case $\beta=2\pi$ so $\mu=0$, and we obtain (note that if we can restore the size of $AdS$ , $R$, by simply multiplying by $R^2$)

\begin{equation}
S = \frac{ \frac{R^2}{ \alpha'}}{ (d-1) }
\end{equation}
The evaluation of the partition function for $r_0=1$ gives the expectation value of the
circular Wilson loop and it is
\begin{equation}
\log Z =  { R^2 \over \alpha'  }
\end{equation}

Note that in ${\cal N}=4$, $\dfrac{ R^2}{\alpha'  }=\sqrt{\lambda} $, while in ABJM  $\dfrac{ R^2 }{ \alpha'  } =\sqrt{2 \pi^2 \lambda} $.

\section{ The precise definition of entanglement entropy}
\label{sec:wald}

When we compute the entanglement entropy using the replica trick we encounter a singular cone.
This cone can be regularized in a variety of ways. One would be to put a boundary condition
at a distance $\epsilon$ from the tip. With this prescription it is very clear that we are computing
the trace of powers of a density matrix. We will call this the ``Hard Wall' prescription.
 Another possibility is to smooth out the cone.
This is the natural computation if we view the entanglement computation as a one loop correction
to a gravitational entropy computation. We call this the ``smooth cone'' prescription. See figure \ref{figure:Cones}.
 The main point we want to stress is that these two
prescriptions can differ by finite terms. Of course the divergent terms are not universal, so we
do not care about divergent terms.
In particular, if we compute the difference between the vacuum entanglement with the entanglement
in the presence of the Wilson loop, then all divergences shoud cancel. But we can nevertheless have a
finite difference. Essentially the same issue is discussed in  \cite{Kabat:1995eq,Solodukhin:2011gn,Donnelly:2012st,Faulkner:2013ana}.

\begin{figure}[h!]
\begin{center}
\vspace{5mm}
\includegraphics[scale=0.35]{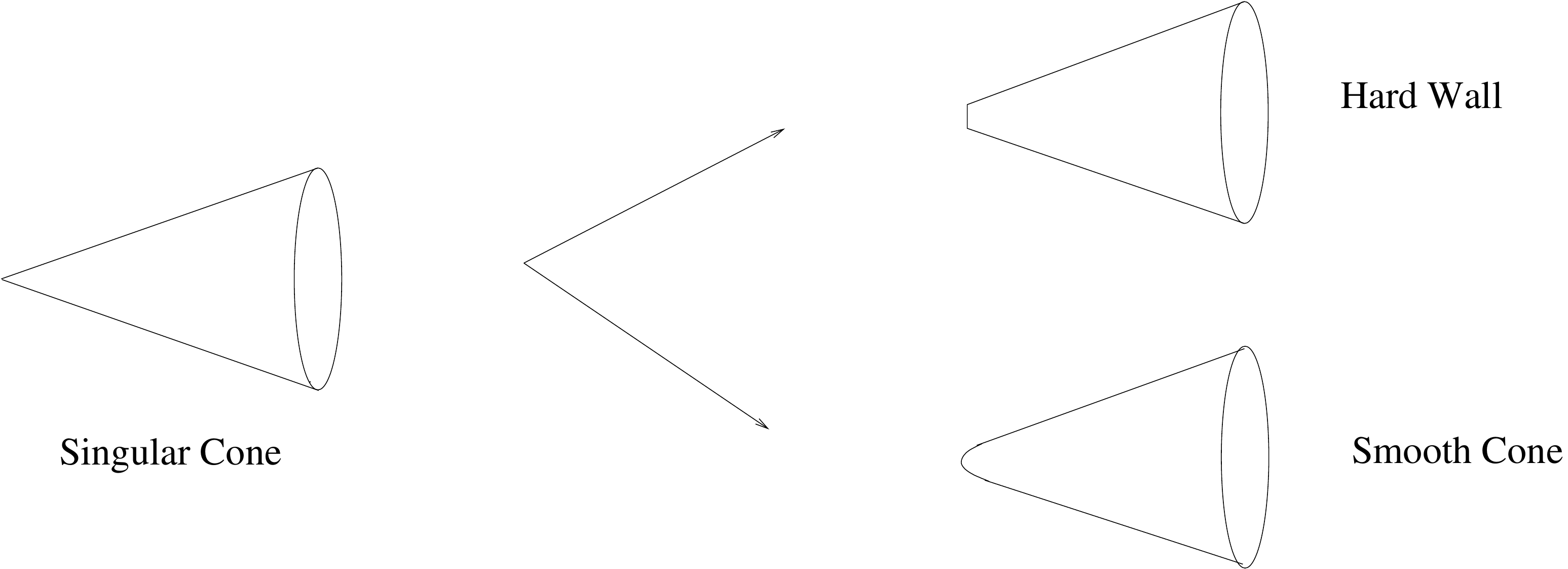}
\vspace{5mm}
\caption{ The singular cone that appears in the replica trick can be regularized in two 
alternative ways. }
\label{figure:Cones}
\end{center}
\end{figure}

This finite difference is due to the presence of the following term in the action
\be
\label{ConfCoupling}
- \int { 1 \over 12} R \phi^2
\ee
that is present for a conformally coupled  scalar field, $\phi$. When we smooth out
the cone we get a finite contribution from the curvature term which is proportional to $n-1$ and is
localized at the tip of the cone. In other words, it is localized on the entangling surface.
Of course, if $\phi$ were a constant this would be the usual area formula arising in
 gravitational entropy. In fact, since $\phi$ is a general function we call this a
Wald term, though the term is usually reserved for higher derivative corrections to the gravitational
action.

Notice that when we map the problem to a thermal problem in the hyperboloid these two prescriptions
translate into either: ``Hard Wall", which is
 changing the temperature everywhere but putting a hard wall at a large distance. Or the ``smooth cone" which would raise the temperature of the hyperboloid for $\rho < \rho_{max}$ but at
larger values of $\rho$ we would revert to $\beta = 2 \pi$.

When we used conformal methods and the insertion of the stress tensor we ignored any possible contribution
from large $\rho$ in hyperbolic space. With the smooth cone prescription this is correct because the
metric reverts to the original metric far away. On the other hand, if we use the ``Hard Wall'' prescription
we can have some boundary terms at the wall. The origin of these
boundary terms is clear. When we vary the metric we need to integrate by parts the variation of
the curvature in \nref{ConfCoupling}, which can give rise to an extra boundary term. This boundary
term is the same as the one giving rise to the area term in the black hole entropy.

The final result is that the difference between the two entropies is
\be \label{difference}
S_{\rm Smooth ~Cone} - S_{\rm Hard~Wall} = \langle  S_{Wald} \rangle  = - { 4 \pi \over 12} \int dA \langle
\phi^2 \rangle
\ee
Here $S_{Wald}$ is the gravitational entropy that we compute from the action \nref{ConfCoupling}.

If we now consider the circular Wilson loop in flat space and view this as the entanglement entropy
of one of the Rindler regions, then this difference is also the same as the difference between
consdering the conformal stress tensor, or a naive stress tensor where we set $R=0$ before we
take the derivative with respect to the metric to compute the stress tensor.

 \section{Free fields: the details}

\label{sec:Aweak}

The free loop $W=e^{ e \int O}$  expectation value is just
\begin{equation}
\log W=\frac{e^2}{2} \int d\tau d\tau' \langle O(\tau) O(\tau') \rangle
\end{equation}
 \subsection{Scalar}
If we set one point at $\rho=0,\theta=0,\phi=0$ and the other point at an arbitrary place, the propagator in $H^3 \times S^1_{2\pi}$ is
\begin{equation}
G(\tau-\tau',\rho,\phi,\theta)=\frac{1}{8 \pi^2} \frac{1}{\cosh \rho-\cos (\tau-\tau')}
\end{equation}

From here sum over the images for going from $\beta=2 \pi$ to  $\beta=2 \pi n$. To do so, one should to do it $m=\frac{1}{n}$ integer and then can analytically continue \cite{Cardy:2013nua}
\begin{equation}
\sum_{k=0}^{m-1} G_{2\pi} (\rho,\tau+\frac{2 \pi k}{m})=\frac{1}{8 \pi^2} \frac{\sinh{m \rho}}{\sinh{\rho}} \frac{m}{\cos {m \tau}-\cosh{m \rho}}
\end{equation}
So we obtain ($\tau \sim \tau + 2\pi n$):
\begin{equation} \label{scalprop}
G_{n}=\frac{1}{8 \pi^2} \frac{\sinh{ \rho/n}}{\sinh{\rho}} \frac{n^{-1}}{\cos {\tau/n}-\cosh{\rho/n}}
\end{equation}

Another useful quantity is the one point function of the scalar field in the presence of the loop
\begin{equation}
\langle \phi\rangle_W= e n \int_0^{2 \pi} d\sigma' G_n (\sigma-\sigma',\rho)=\frac{e}{4 \pi \sinh \rho}
\end{equation}

So, $\langle \phi^2  \rangle_W= \dfrac{e^2}{16 \pi^2} \dfrac{1}{\sinh^2 \rho}$.

 \subsection{Gauge field}
 The Maxwell equation in Feynman gauge is
 \begin{equation}
 -g_{\mu \nu} \nabla^2_{H_3 \times S^1} +R_{\mu \nu}=-g_{\mu \nu} \nabla^2_{H_3 \times S^1} -2 \delta_{\mu \nu}^{i j} g_{\mu \nu}
 \end{equation}
 Where $i,j$ denote the directions in the hyperboloid. The solution to this is $\langle A_{\mu} A_{\nu} \rangle = g_{\mu \nu} D_{m^2}$. That is, we have the propagator of a massless scalar for the temporal modes and the propagator of a $m^2=-2$ scalar for the spatial directions.  We have the explicit propagator for a massive scalar field in $H_3 \times R$, \cite{Camporesi:1990wm}
 \begin{equation}
 G^{\infty}_{m^2}=\frac{\sqrt{m^2+1}}{4 \pi^2 } \frac{\rho}{ \sinh \rho \sqrt{\rho ^2+t^2}}  K_1 \left( \sqrt{m^2+1} \sqrt{t^2+\rho ^2}\right)
 \end{equation}
 In \cite{Camporesi:1990wm}, they analytically continued it from the sphere and they got $H_1^{(2)}$ instead of the Bessel function. Demanding  regularity when $t \rightarrow \pm \infty$ requires that we pick $K_1$.
 This gives us the conformally coupled scalar propagator for $m^2=-1$, with the proper normalization. This agrees with the
 $n\to \infty $ limit of \nref{scalprop}

From here we can obtain the solution at finite temperature
$G_{\beta}=\sum_{n=-\infty}^{\infty} G_{\infty}(t+n \beta)$.
So the integral $\log W_{\beta} \propto \beta \int_0^{\beta} d \tau G_{\beta}(\tau)=\beta \int_{-\infty}^{\infty} dt G_{\infty}(t)$.

\subsection{The need for a boundary term in  hyperboloid}

In section \ref{sec:n4check} , we saw that for the scalar field $\partial_{\beta} \log W= -\int \langle T_{\tau \tau}^N \rangle=0$ (by $T^N$ we mean the stress tensor without adding the improvement term). However if we  explicitly plug the stress energy tensor and the time independent solution $\phi_w=\sinh^{-1} \rho$, we obtain
\begin{equation}
-\int d \rho T^N_{\tau \tau}=-\int_{\epsilon}^{\infty} d\rho ((\partial_\rho \phi_w)^2-\phi_w^2)=1
\end{equation}

This does not seem to be compatible with the statement $\partial_{\beta} \log W= -\int \langle T_{\tau \tau}^N \rangle$. The reason is that we have to impose further boundary conditions.
The time independent solutions to the scalar equation are $\phi=\frac{a+b\rho}{\sinh \rho}$. We want to pick the solution with $b=0$. To do that we fix the boundary conditions at infinity: $\sinh \rho(\phi+\partial_{\rho} \phi))|_{\rho=\infty}=0$.  Of course the same equation is true for the variation. So we should impose the boundary condition in the variation of the action.

The action will apply this boundary condition  if we include a boundary term:
$S=S-\frac{1}{2} \int_{\rho=\infty} \sinh^2 \rho \phi \partial_{\rho} \phi$.
This fixes everything:
\begin{equation}
-\int d\rho T^N_{\tau \tau}=-\int_{\epsilon}^{\infty} d\rho \sinh^2 \rho ( (\partial_\rho \phi_w)^2-\phi_w^2) + \int_{\rho=\infty} \sinh^2 \rho \phi_w \partial_{\rho} \phi_w=0
\end{equation}

\section{Comments about the 1/2 BPS loop}
\label{sec:ABJMw}

 The half BPS Wilson loop \cite{Drukker:2008zx2}  is   more complicated. It is formulated in terms of the holonomy of a superconnection. This superconnection is a matrix whose diagonal elements look like those of the 1/6 BPS loop and has fermions in the off-diagonal terms. For the three sphere, it turns out that it is in the same cohomology as the sum of the two 1/6 BPS loops, so we can easily compute it from the matrix model integral \nref{ABJMMM}

 \begin{equation}
 \langle W^{1/2} \rangle_{S^3}=\langle W^{1/6} \rangle_{S^3}+\langle \bar{W}^{1/6} \rangle_{S^3}
 \end{equation}

 Where $\langle \bar{W}^{1/6}(\lambda)\rangle_{S^3}=\langle W^{1/6}(-\lambda)\rangle_{S^3}$ is just the 1/6 loop for the second $U(N)$. In the planar limit, $\langle W^{1/2} \rangle$ has a very simple expression in terms of $\kappa$
  (see eqn. \nref{kappa} for the relation between $\kappa$ and $\lambda$)  \cite{Drukker:2010nc}
\begin{equation}
 \langle W^{1/2} \rangle=\frac{\kappa}{8 \pi \lambda  }
\end{equation}

 Note that it is real.  For the winding loop there is a small change \cite{Klemm:2012ii} $ \langle W_m^{1/2} \rangle_{S^3,MM}=\langle W_m^{1/6} \rangle_{S^3}-(-1)^m \langle \bar{W}^{1/6} \rangle_{S^3}$.

 Now we would like to find a simple expression for the loop in the squashed sphere, at least near $b \sim 1$. The arguments that it is in the same cohomology as the sum of the two 1/6 BPS loop seems to also apply here. Very roughly, the localization locus is $\sigma=$ const and all the other fields are set to zero, so we expect that it is just the sum of the two loops.
 To first order in $b-1$ we expect that we should simply be computing the expectation value of $ e^{ b \mu_1 } + e^{ b \nu_1}$
 in the matrix model with $b=1$. Here $\mu_i$ and $\nu_i$ are the eigenvalues of the matrix model for each of the two groups.

This would lead to
\be \label{Firstpos}
\partial_b \langle W^{1/2}(S^3_b) \rangle|_{b=1}  =
\left.  \partial_m \left( \langle W_m^{1/6} \rangle_{S^3}+ \langle \bar{W}^{1/6}_{m}\rangle _{S^3} \right) \right|_{m=1}
\ee
However, one can check that by expanding \nref{ABJMMM}in powers of $\lambda$ that this expression vanishes (at $m=1$). This fact can balso
be checked numerically and at strong coupling.
This seems incorrect since this expression is also determining the expectation value of the energy in the presence of the
Wilson loop. At strong coupling the energy seems to be non-zero. We have not understood how to interpret this result.

On the other hand, if we were to  assume that the right answer is obtained by taking the derivative with respect to
$m$ of the multiply wound Wilson loop, then we would obtain instead
\be \label{Secondpos}
\partial_b \langle W^{1/2}(S^3_b) \rangle|_{b=1}  = \left.
\partial_m (\langle W_m^{1/6} \rangle_{S^3}-(-1)^m \langle \bar{W}_{m}^{1/6} \rangle_{S^3}) \right|_{m=1}
\ee
  This result is non-zero only due to the derivative of the $(-1)^m$ term, given that \nref{Firstpos} is zero.
However, this result,  $\text{Re} (i \bar{W}^{1/6})$,
 is an odd function of $\lambda$, so both the corresponding $B$ function and the entropy will not be invariant under parity.
 It is strange that if $k<0$ we have negative Bremmstrahlung.
 However, this result is  in agreement with the perturbative two loop computation in \cite{Griguolo:2012iq}.
The one loop correction at strong coupling was computed in \cite{Forini:2012bb}, but it is not in
agreement. However, this could be due to some subtleties in the precise relation between the radius of AdS
and $\lambda$.

 In summary, the situation with the 1/2 BPS Wilson loop is confusing and requires more thought.

 The recent paper \cite{Marmiroli:2013nza} obtains a result for the Bremsstahlung function for
 1/2 BPS Wilson loop which is even in $\lambda$.

\end{document}